**Evaluation of the suitable analytical techniques for the investigation of the toxic elements and compounds in the Pyrotechnic materials (Green crackers)**

Darpan Dubey[1], Rohit Kumar[2†], Abhishek Dwivedi[1], Rahul Agarwal[3], Awadhesh Kumar Rai[1*]

[1]Laser Spectroscopy Research Laboratory, Department of Physics, University of Allahabad, India -
211002

[2]Department of Physics, CMP College, University of Allahabad, India-211002

[3]Centre of Food Technology, University of Allahabad, India-211002

Corresponding author Email: awadheshkrai@gmail.com

**Abstract:**

The present manuscript reports the elemental as well as molecular study of the Green Crackers. Laser-induced breakdown spectroscopy has been used for elemental analysis, UV-Vis and Photoacoustic Spectroscopy (PAS) are used for molecular study of the green crackers. The spectral lines of several elements including heavy/toxic such as Al, Ba, Sr, Cr, Cu are observed in the LIBS spectra of green crackers like present in normal crackers. In addition to this, the electronic bands of diatomic molecules like AlO, SrO, and CaO are also observed in LIB spectra of the green crackers. PAS, which is non-destructive, useful for scattering & opaque substances, is more suitable than the UV-VIS method for the investigation of the various organic compounds/molecules present in the firecrackers. Molecular bands of these molecules (AlO, SrO and CaO) are also in the absorption spectra of the crackers recorded using PAS technique and UV-Vis spectroscopy technique. In addition to these, absorption bands of some additional compounds/molecules like AlO, SrO, $CaCO_3$, $KNO_3$, $NH_4NO_3$, $NHClO_4$ are also observed in the PA spectra of the green crackers, which show that PAS is more appropriate technique than the UV-VIS method for the investigation of the organic compounds/molecules in firecrackers. To determine the exact concentration of the constituents (Al, Cr, Cu) in green crackers AAS has been used. The results of the present manuscript show that the green crackers are also toxic for the environment as well as for humans although with lesser intensity than traditional/normal crackers.

## 1. Introduction:

In India, Diwali, Dusshera, and many other festivals are celebrated with full of enthusiasm like a new year all over the world. Independence days of the countries, republic days and all the special days like opening ceremony and finale day of shows are celebrated in all over the world. The crackers play a major contribution to adding the happiness and joy of people in celebrating these functions. Sky crackers, Bombs, Ground crackers, Sparklers, Spinner are widely used during celebrations. (Awasthi et al., 2017; Report, 2020; Russel). The crackers emit a lot of fumes containing pollutants, which might be toxic. This fume contains toxic gases in molecular form Which pollute the environment, human beings as well as all the creators of the earth (Martín-Alberca & García-Ruiz, 2014; R.B. Soni 1; Udaya Prakash et al., 2019). These toxic gases emitted from the traditional crackers causes a lot of fatal disease for human. These gases are harmful not only to humans, but these toxic gases are also harmful to other living organisms as well. The gas contains different molecules of sulphate and nitrate, these sulphate and nitrate break into oxides of sulphur and nitrogen, responsible for acid rain (Martín-Alberca & García-Ruiz, 2014). The breaking of $SO_2$ and $NO_2$ lead to acid rain. Acid rain would become a big danger for the historical buildings and forts. Nitrogen dioxide and nitric oxide and sulphur dioxide ($NO_2$, NO, SO2) are highly toxic by inhalation (Martín-Alberca & García-Ruiz, 2014; R.B. Soni 1; Udaya Prakash et al., 2019). Additionally, the composition used in crackers is gunpowder/black powder it works as a propellant in crackers. Gunpowder is generally made by $KNO_3$ and Burning of gunpowder/black powder is leads to low blood pressure problems and the reaction of burning is given by the following equation-

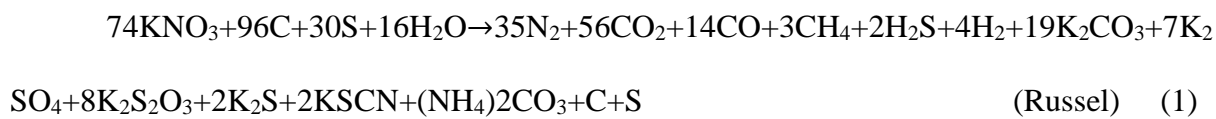

$$74KNO_3+96C+30S+16H_2O\rightarrow35N_2+56CO_2+14CO+3CH_4+2H_2S+4H_2+19K_2CO_3+7K_2$$

$$SO_4+8K_2S_2O_3+2K_2S+2KSCN+(NH_4)2CO_3+C+S \qquad\qquad (Russel) \quad (1)$$

The above reaction contains a composition containing saltpeter (75.75%), Charcoal (11.7%), Sulphur (9.7%), and moisture (2.9%) which clarifies that the burning of crackers releases toxic gases (Russel). As the temperature of Gunpowder/black powder is increasing the decomposition of the black powder set in progress

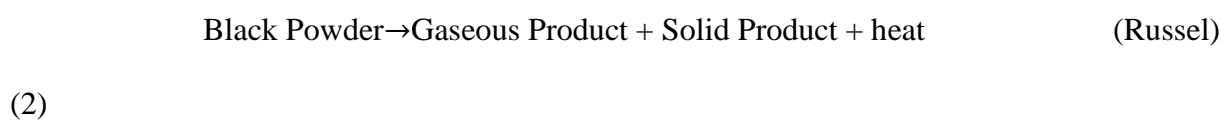

$$Black\ Powder\rightarrow Gaseous\ Product + Solid\ Product + heat \qquad\qquad (Russel)$$
(2)

Apart from propellant to produce hot gases, Firecrackers also contains different kind of elements like Al, Ba, Ca, Cu, Sb, Sr, and K to produce a different kind of exciting colours. The beauty of these elements is that they produce exciting colours but have their own complexity. The side effect of Potassium compound (K) can contaminate ground and surface water, become a cause of thyroid problems in humans and animals (Awasthi et al., 2017; Report, 2020). The compound of (Ba) can produce the green colour and is responsible for problems related to respiratory, muscular weakness. Copper compound (Cu) led to cancer, harmonic imbalance, skin condition, and accumulation with the body. The compound of Al produces white colour and is responsible for skin problems, Alzheimer's disease, and accumulation within the body. The compound of Antimony (Sb), which is used as colouring agent, affects respiratory irritation and leads to lung cancer.

Residues which remain after the burning of firecrackers can be hazardous immediately or after some time and reactions with other elements in the atmosphere are also found dangerous specially for pregnant women. For mentally disturbed/ill people, it leads to depression, fear, and stress (Awasthi et al., 2017; Report, 2020). These elements are mixed in firecrackers to produce different colors in a firework. Al, Ti, and Mg produce silver color, Ba is mixed to produce green color, Sr and Li for Red, Cu for Blue, Na for yellow, Fe for golden, Ca for orange color, V is used to produce silver sparks, and Sb is used for glitter effect in crackers. Some elements and their compounds in fireworks are used as an oxidizer such as Al, O, Mn, K, fuel. The nickel acts as an electric firing device for firecrackers. Among several oxidizers like Lead Dioxide/Nitrate/Chloride and Perchlorate (Ammonium and Potassium), chlorine is widely used in firecrackers (Awasthi et al., 2017; Report, 2020). The role of oxidizers in crackers is to produce excess oxygen to make a better explosion.

Thus it is of utmost necessity to manufacture/develop crackers having less concentration of toxic constituents, that have reduced sound and intensity compared to normal/traditional crackers which reduced the accidents in society. To achieve the goal, the Council of Scientific Indian Research (CSIR), India, a premier national R&D organization, and National Environmental Engineering Research Institute (NEERI) jointly have developed crackers that get rid of all shortcomings of traditional and normal crackers by minimizing or degrading the concentration of toxic constituents up to 30 % in a special type of cracker called "green crackers" [7]. These crackers are eco-friendly because these crackers pose a lesser impact on the environment and have a lesser health risk. The raw materials used in green crackers are less polluting and the particle emission into the atmosphere is reduced by suppressing the dust. Gases from the burning of firecrackers are harmful too.

Our objective in the present manuscript is to investigate the presence of toxic elements/compounds in green crackers and compare their harmfulness to human health and environmental pollution with normal/traditional crackers.

The particulate matter (PM) level is an essential factor that needs to be studied. Lin et al (Lin, 2016) and Prakash et al (Prakash et al., 2019) have focused their study on the change in the particulate matter (PM) level after burning the crackers during Diwali. The burning of crackers generates a dense smoke cloud containing many pollutants particles of size like $PM_{2.5}$ (measured on scale of 2.5 nm) and $PM_{10}$ (measured on scale of 10 nm), which enhance the total suspended particulate matter (TSP) (Lin, 2016). In Diwali evenings, some molecules like $NO_2$ are increased 2-3 times and $SO_2$ is increased nearly ten times compared to non-Diwali day evenings in some parts of India (Prakash et al., 2019). It is also reported that the level of charcoal and concentration of some toxic elements like Ba, K, Al, and Sr have increased abruptly during the festive season (Prakash et al., 2019). The rise in $PM_{2.5}$ levels has also been recorded in other countries like the USA and China (Alenfelt, 2000; Martín-Alberca & García-Ruiz, 2014). Much work has already been done on increasing PM levels after burning fireworks and their hazards (Lei et al., 2014).

For the elemental detection of crackers, there are numerous conventional techniques like Inductively coupled plasma optical electron spectroscopy (ICP-OES), Inductively coupled plasma mass spectroscopy (ICP-MS), and X-ray Fluorescence (XRF). These conventional techniques are either time-consuming, have rigorous sample preparation methods, or are unable to detect the lighter elements. Therefore, we need a quick, reliable, and requires almost no sample preparation technique that can detect and quantify all the elements including lighter elements and heavy/toxic elements simultaneously (Awasthi et al., 2016). LIBS is a robust and fast technique, that can be used for the detection of all types of constituents present in any type of material sample.

Therefore in the present manuscript LIBS combined with Chemometric analysis has been used to detect and determine the presence of toxic elements in green crackers. For the classification/prediction of characteristics of different crackers, the multivariate technique, here Principal Component Analysis (PCA) is used. The main objective of applying multivariate analysis is to minimize the spectral data's dimension with lesser factors to describe data. Chemometric techniques generally approve with LIBS data's addition to upgrade the ability of discrimination, regression, and prediction (Awasthi et al., 2017; Martín-Alberca & García-Ruiz, 2014). In this paper, Principal Component Analysis (PCA) is applied to the LIBS spectra of green crackers for the classification of samples based on spectral features of the green cracker's samples (Awasthi et al., 2017).

## 2. Materials and Methods:

Six green crackers, details of the samples are given in Table 1, are procured from the local market of Delhi, India.

**Table 1:** Detail of the green cracker samples and their weight

| S.N. | Green cracker Sample | Type | Amount of sample taken |
|------|----------------------|------|------------------------|
| 1 | $G_1$ | Sparkler | .706g |
| 2 | $G_2$ | Sparkler | .706g |
| 3 | $G_3$ | Sparkler | .706g |
| 4 | $G_4$ | Flower Pot | .706g |
| 5 | $G_5$ | Flower Pot | .706g |
| 6 | $G_6$ | Flower Pot | .706g |

To record the LIBS spectra of green crackers, the powder was extracted and crushed gently to get the homogeneous grain with the help of mortar and pestle. The crushed samples were sieved with the 150 μ sieve and made pellet of amount 0.706g with the help of KBr pellet machine (H-Br Press MODEL M-15) having pressure 1200 Pascal.

To get the optimum spectral intensity i.e. to get the maximum signal to background ratio and signal to noise ratio, two experimental parameters, laser pulse energy and gate delay were optimized. Consequently, a gate delay and gate width of 1 µs and 5 µs, respectively, were seen as appropriate for a better LIBS signal. LIBS spectra of all the samples were recorded at these optimized experimental conditions.

A Q-switched, pulsed Nd: YAG laser (Continuum Surelite III-10, pulse width FWHM = 4 ns at 532 nm, maximum deliverable laser energy 425 mJ per pulse, variable repetition rate 1–10 Hz) is used as an excitation source. A plano-convex lens (f = 15) is used to focus the laser beam onto the sample in the ambient air, which produces the laser-induced plasma. The sample is mounted onto a translational sample stand so that each laser pulse is incident at a fresh spot on the sample surface. This avoids the formation of deep craters and ensures reproducibility on a shot-to-shot basis. The radiation emitted from the plasma was collected by a collection optics assembly (CC 52, Andor), which was aligned at a 45-degree angle to the incident laser beam. Further, the emission light was directed towards the spectrometer (Andor, Mechelle 5000) equipped with an ICCD camera (Andor, iSTAR DH734) through an optical fiber of core diameter 600 µm. Shematic diagram of laser is shown in Figure 1.

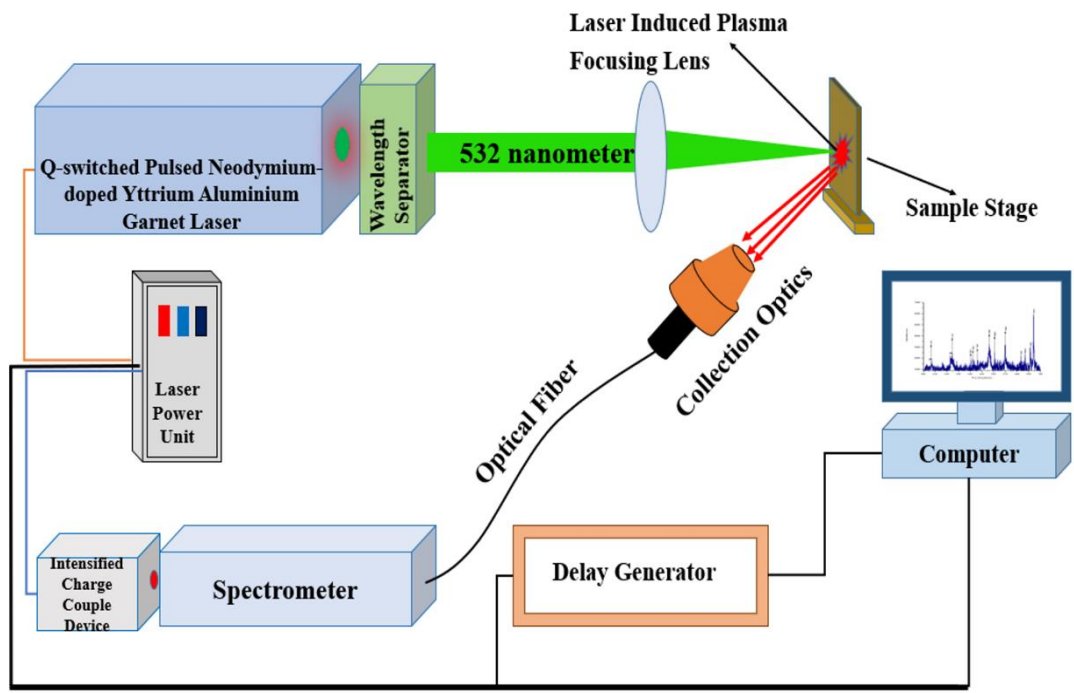

**Figure 1**: Shematic diagram of LIBS set-up

For the identification of molecular band PA spectra of the samples were recorded in the range of 250-850 nm using experimental set-up which consist a xenon arc lamp (S/N 1040 HORIBA Canada, OBB Corporation), chopper (model SR-540, Stanford, USA), PA cell (contains a sample cup) and a lock-in amplifier (model SR-830, Stanford, USA). Sample is used in powder form. Block diagram of photoacoustic set-up is shown in Figure 2.

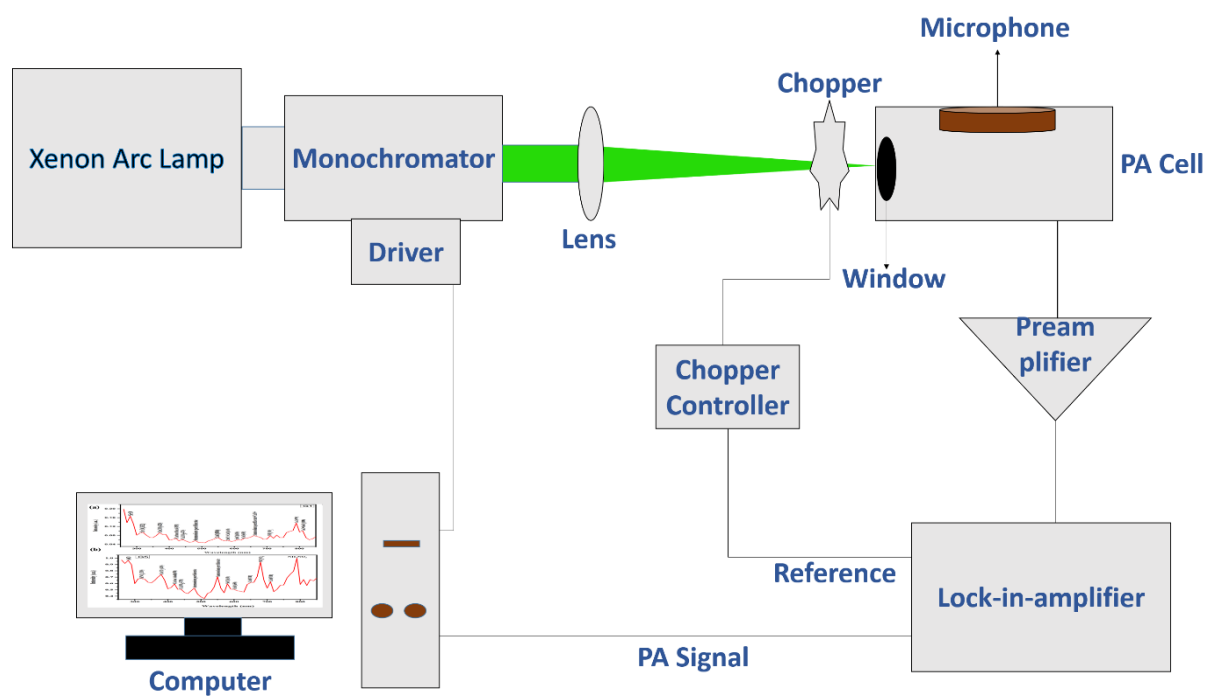

**Figure 2**: Shematic diagram Photoacoustic set-up

UV-Vis spectra were recorded in the range of 220-850 nm using an Unicam-5625 spectrometer in absorbance mode. The sample is dissolved in distilled water for the measurement at room rempreture.

## 2.1. Preparation of samples and standard for Atomic Absorption Spectroscopy (AAS):

For AAS analysis 0.25 gm of the sample was weighed and taken in a digestion tube after this, it was treated with concentrated nitric acid ($HNO_3$). Now blank sample was prepared by applying 1 ml of nitric acid into an empty digestion flask. Now the flask was warmed up to 2 hours into a digestor at 450 ℃ (pelican equipment, Chennai, India) until a transparent solution was obtained. When digestion was completed, a few drops of concentrated Hydrochloric acid (HCl) were added to it. Now solution was heated slowly in low flame. Now the residue was collected and subjected to digestion and filtrate was collected. When the solution was Cool down, the solution was filtered with Wattman paper No.42. Now, this filter was taken into 1000 ml of volumetric flask and was made by adding the distilled deionized water (Agrawal, 2011). The slandered solution of Chromium (Cr), Copper (Cu), Aluminium (Al) were prepared from the stock standard solutions containing 500 ppm of the element in the normal concentration of nitric acid. In Calibration and measurement of green firecrackers sample, Atomic Absorption Spectrometer (Analyst 700, Perkin Elmer, USA). Using linear correlation, the calibration curve was plotted for each constituent individually. In between measuring/calculating of concentration of each constituent, the necessary correction was made after taking the blank reading.

## 3. Result and Discussion:

### 3.1. Qualitative Analysis:

All six types of green crackers are analyzed with the help of the LIBS technique using the National Atomic Spectroscopy Database (NIST) (Kramida et al., 2015) and Chemical Spectroscopy by W.R. Brode (Brode, 1958).

The LIBS spectra of the six green crackers shown in Figure 1 contain the spectral lines of several elements including some toxic elements. The list of elements observed in the spectra of firecrackers is listed in Table 2 The molecular bands observed in LIBS spectra of different samples are also listed in Table 2.

**Table 2**: All the elements and molecules present in green crackers samples

| Sr. No | Sample | Elements | Molecular Bands |
|---|---|---|---|
| 1 | $G_1$ | C, Al, Mg, Ca, Sr, K, O, N, Na, Cu, Cr, Mg | CaO, SrO |
| 2 | $G_2$ | C, Al, Mg, Ca, Sr, K, O, N, Na, Cu, Cr, Mg | CaO, SrO |
| 3 | $G_3$ | C, Al, Mg, Ca, Sr, K, O, N, Na, Cu, Cr, Mg | CaO, SrO |
| 4 | $G_4$ | Al, Mg, H, N, O, K, Na, Ca, Si, Cu, Cr, Sr | AlO |
| 5 | $G_5$ | Al, Mg, H, N, O, K, Na, Ca, Si, Cu, Cr, Sr | AlO |
| 6 | $G_6$ | Al, Mg, H, N, O, K, Na, Ca, Si, Cu, Cr, Sr | AlO |

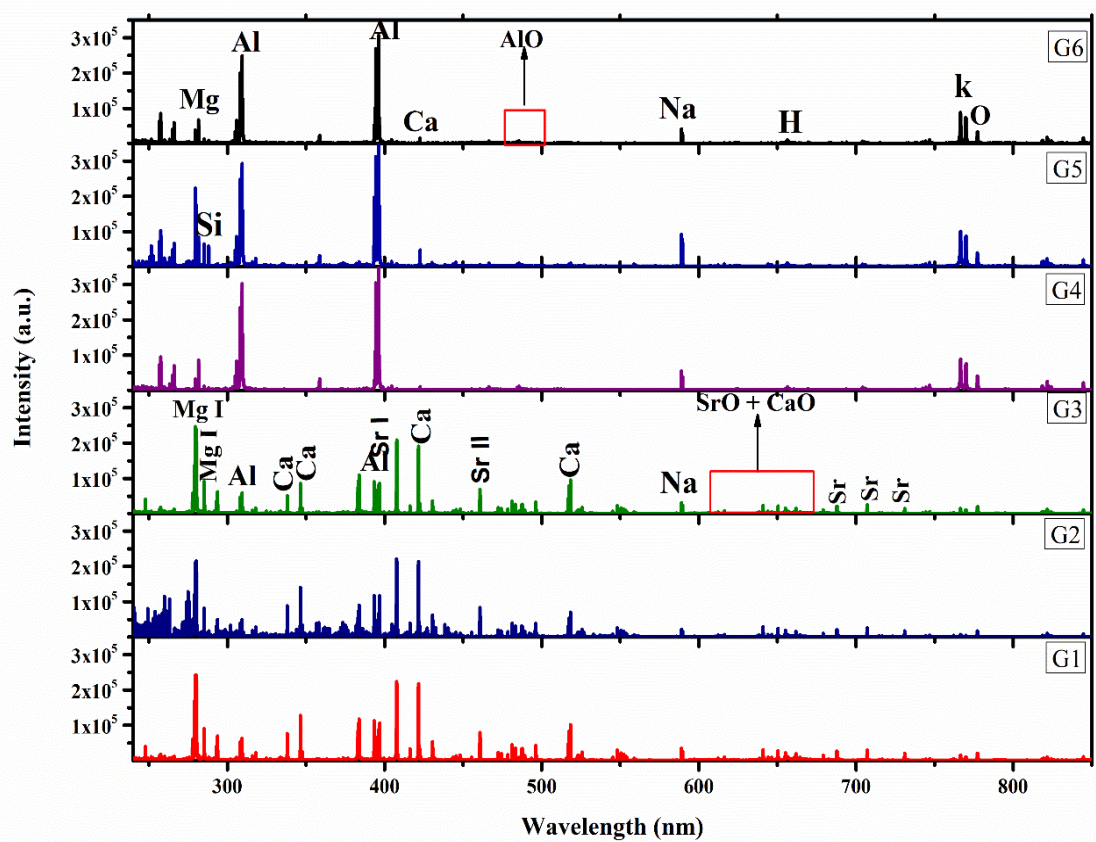

**Figure 3**: Stack plot of LIBS spectra of all six types of green crackers

After analyzing the LIBS shown in Figure 3, we can divide these green crackers into two groups. The first group contains $G_1$, $G_2$, and $G_3$ as the same spectral line of the constituents like C, Al, Mg, Ca, Sr, Na, K, O, and N and molecular bands of CaO corresponding $\Delta v=0,+1$, SrO, and SrCl are present in all three groups. $G_4$, $G_5$, and $G_6$ samples fall in the second group because this group of samples containing the same spectral lines of the constituents like Al, Mg, Ca, Na, H, and K. In addition to this second group of samples contains the molecular band of AlO corresponding $\Delta v=0, -1,+1$ because of more availability of the Al in this group samples as compared to the samples $G_1$, $G_2$, and $G_3$ of the group first.

## 3.2. Semiquantitative Analysis:

The intensity of the spectral lines of the elements are used to determine their concentration in the target material, then the laser-induced plasma should satisfy the following three criteria **i)** plasma should be Stochiometric **ii)** plasma should be optically thin and **iii)** plasma shoul be in Local Thermal Equilibrium. In the present experiment the above three criteria are satisfied and given in the following paragraph:

### 3.2.a. Stochiometric ablation:

If the plasma is a true representative of the sample's composition, then the Laser-induced plasma can be considered stoichiometric. It has been reported that if the irradiance of Laser on the surface of the sample is higher than $10^9$ W/cm$^2$, then the material acquires its vaporization temperature due to that surface of the sample explored before the surface layer can vaporize this condition shows the stoichiometric ablation (Singh Maurya et al., 2014). In the present experiment, the laser energy is 50 mJ per pulse. The focal spot diameter is 12.70 μm (the focal spot was measured using D = 4λf/πd, where the laser wavelength is λ (532 nm), the focal length of the lens used is f (15 cm), the laser beam diameter is d (8 mm) (Awasthi et al., 2016). Therefore, the fluence and irradiance of the Laser are calculated to be $4.14\times10^4$

J/cm$^2$ and 3.14×10$^{12}$ W/cm$^2$ respectively. Thus, it is clear that Laser-induced plasma satisfies the stoichiometric ablation as the Laser irradiance is higher than 10$^9$ W/cm$^2$.

### 3.2.b. Optically thin plasma:

Sometimes, the emitted photons inside the plasma are absorbed by another atom in a lower energy state; thus, in the resulting spectrum, the intensity of that particular photon is reduced, known as self-absorption. It occurs when the plasma is optically thick. Thus, to avoid such a situation, Laser-induced plasma necessarily should be optically thin. The laser-induced plasma is supposed to be optically thin if the intensity ratio of two interference-free emission lines of the same element with nearly the same upper energy level is equal to the ratio of the multiplication of the transition probability, statistical weight, and inverse of the wavelength of these spectral lines. The intensity ratio of Al in the different samples was evaluated and it is tabulated in Table 3. It is clear from Table 3 that these intensity ratios are nearly equal to the theoretical ratios (Singh Maurya et al., 2014). Thus the present laser-induced plasma is optically thin.

**Table 3:** Calculation of intensity ratio of the two atomic lines for Aluminum, 308.2/309.2 nanometer in the Laser-induced breakdown spectra of the different samples.

| Samples | $A_{ki}g_k\lambda'/A'_{ki}g'_k\lambda$ | Intensity Ratio, $I/I'$ |
|---|---|---|
| G1 | 0.84 | 0.78 |
| G2 | 0.84 | 0.86 |
| G3 | 0.84 | 0.92 |
| G4 | 0.84 | 0.76 |
| G5 | 0.84 | 0.72 |
| G6 | 0.84 | 0.76 |

### 3.2.c. Local Thermal Equilibrium:

A Laser Induced Plasma is said to be in local thermal equilibrium (LTE) if and only if it fulfills the necessary and sufficient conditions. For necessary conditions, the electron density in Laser-induced plasma should be greater than the McWhirter limit. For sufficient conditions, the ionization temperature should be within 15% of the excitation temperature (Singh Maurya et al., 2014). The Boltzmann equation (sometimes referred to as Maxwell-Boltzmann equation) is given below,

$$I_\lambda^{ki} = F C_s A_{ki} g_k \frac{exp\ (-E_k/kT)}{U_s(T)} \qquad (3)$$

where $A_{ki}$ is the transition probability, $E_k$ is the upper energy level, $g_k$ is the statistical weight, F is the experimental factor, $C_s$ is the concentration of the constituents present in the sample, k is Boltzmann constant and $U_s(T)$ is the partition function (Singh Maurya et al., 2014).

Taking the natural log of two sides in the above equation (3), gives

$$ln\frac{I_\lambda^{ki}}{A_{ki}g_k} = -\frac{E_k}{kT} + ln\frac{FC_s}{U_s(T)}$$

$$(4)$$

For the calculation of the experimental factor F, it is assumed that the sum of the concentrations of all species is one, i.e.

$$\sum C_i = 1 \qquad (5)$$

A graph between ln (I/$A_{ki}$.$g_k$) and $E_k$ is plotted (Boltzmann plot) and shown in Figure 4.b). The lower limit of electron density, i.e. McWhirter limit, has been calculated using relation,

$$N_e\ (cm^{-3}) > 1.6 \times 10^{12} T^{1/2} (\Delta E)^3 \qquad (6)$$

Here, the electron density in the laser-induced plasma is $N_e$, T is plasma temperature in Kelvin and the maximum energy difference of a spectral line is ☐E.

Experimentally, the electron density is calculated by using the Full-Width at Half Maxima (FWHM) ($\Delta\lambda$) of the Stark broadened line,

$$N_e = 10^{16} \frac{\Delta\lambda}{2w}$$

(7)

where the electron density is $N_e$ and electron impact parameter is w, whose value can be calculated from plasma spectroscopy, and $\Delta\lambda$ is calculated by using the following relation

$$\Delta\lambda_{True} = \Delta\lambda_{Observed} - \Delta\lambda_{Spectrometer} \tag{8}$$

where $\Delta o_{bserved}$ is Full-Width at Half-Maxima (FWHM) of the stark broadened spectral line (Figure 4.b) and $\Delta s_{pectrometer} = 0.05$ for our experiment. In the present study, the McWhirter limit calculated using equation (6) is $2.39 \times 10^{15}$ cm$^{-3}$ and the value of $N_e$ calculated from equation (7) is $1.6 \times 10^{17}$ cm$^{-3}$, which fulfils the necessary condition for LTE.

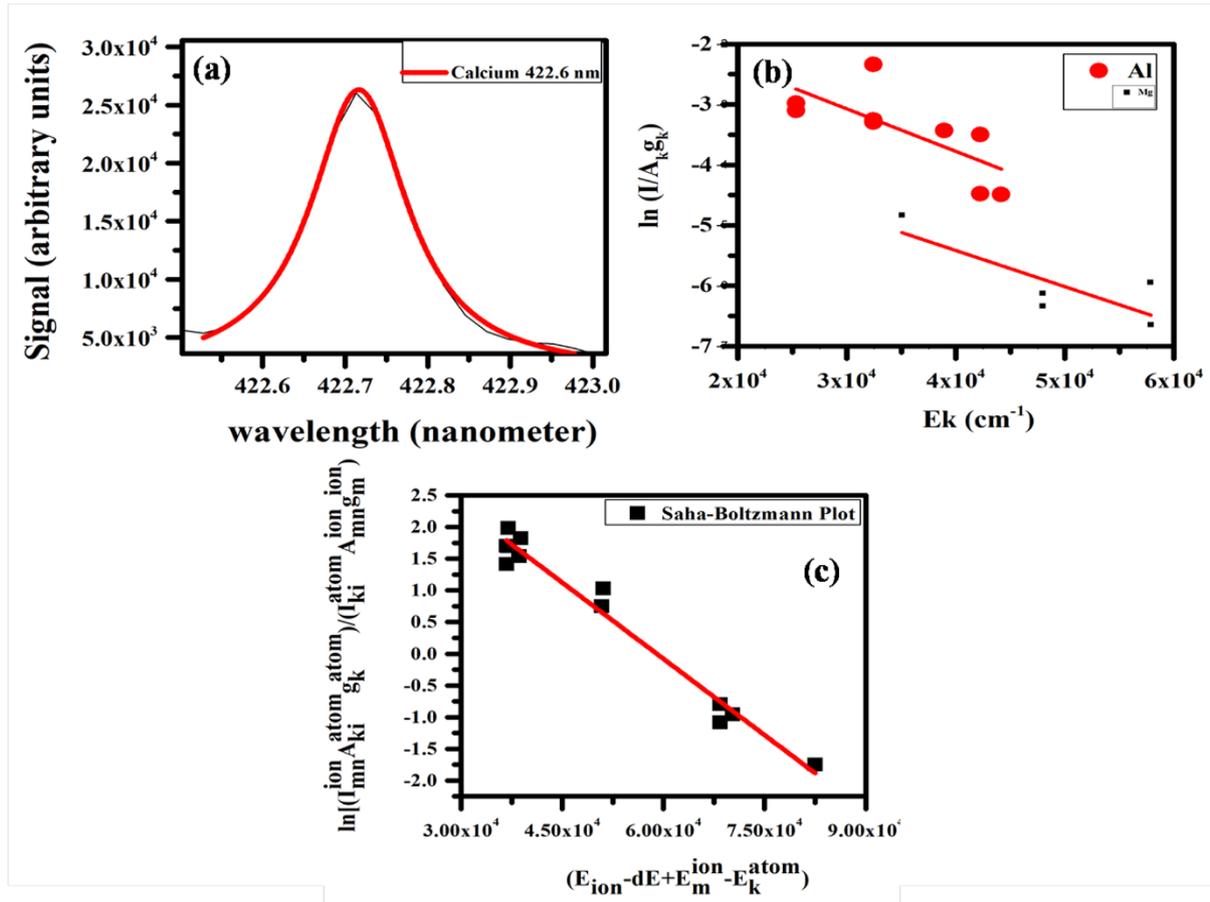

**Figure 4.a):** Lorentzian plot of Calcium at 422.7 nanometres for the sample G1 to determine .**b):** Boltzmann plot for sample G3 to determine plasma temperature for Al and Mg element. **c)**: Saha-Boltzmann plot to determine the ionization temperature and to validate the local thermal equilibrium.

For sufficient condition, ionization temperature has been calculated using the Saha-Eggert equation is given below,

$$ln\frac{I_{mn}^{ion}A_{ki}^{Atom}g_k^{atom}}{I_{ki}^{atom}A_{mn}^{atom}g_m^{ion}} = -\frac{(E_{ion} - dE + E_m^{ion} - E_k^{atom})}{kT} + ln\frac{2(2\pi m_e kT)^{3/2}}{N_e h^3} \qquad ,$$

(9)

where the mass of the electron is $m_e$, Plank constant is h, $E_{ion}$ is the first ionization potential, lowering correction parameter is dE, which is the correction term in the first ionization potential comes due to high pressure in plasma plume; $E_m^{ion}$ and $E_k^{atom}$ are upper energy levels of ionic and atomic lines of the element which have the transition probabilities $A_{ki}^{Atom}$ & $A_{mn}^{atom}$ and statistical weight $g_k^{atom}$ & $g_m^{ion}$ respectively.

A plot between $\ln(\frac{I_{mn}^{ion}A_{ki}^{Atom}g_k^{atom}}{I_{ki}^{atom}A_{mn}^{atom}g_m^{ion}})$ and $(E_{ion} - dE + E_m^{ion} - E_k^{atom})$ is drawn and shown in Figure 4.c). The evaluated value of ionization temperature using the Saha-Eggert equation is 16231±431K, close to the excitation temperature 18329± 817K, with a difference of almost 11%. Thus, the necessary, and sufficient conditions, satisfied and thus LTE holds in the present experiment.

Our Laser-induced plasma satisfies the above three conditions of Stoichiometric, optically thin plasma, and Local thermal equilibrium. This implies that LIBS spectra are useful for Quantitative analysis. Here, the Intensity of the spectral lines of the Constituents (Al, Cu, Cr and Sr) are evaluated and plotted against the samples.

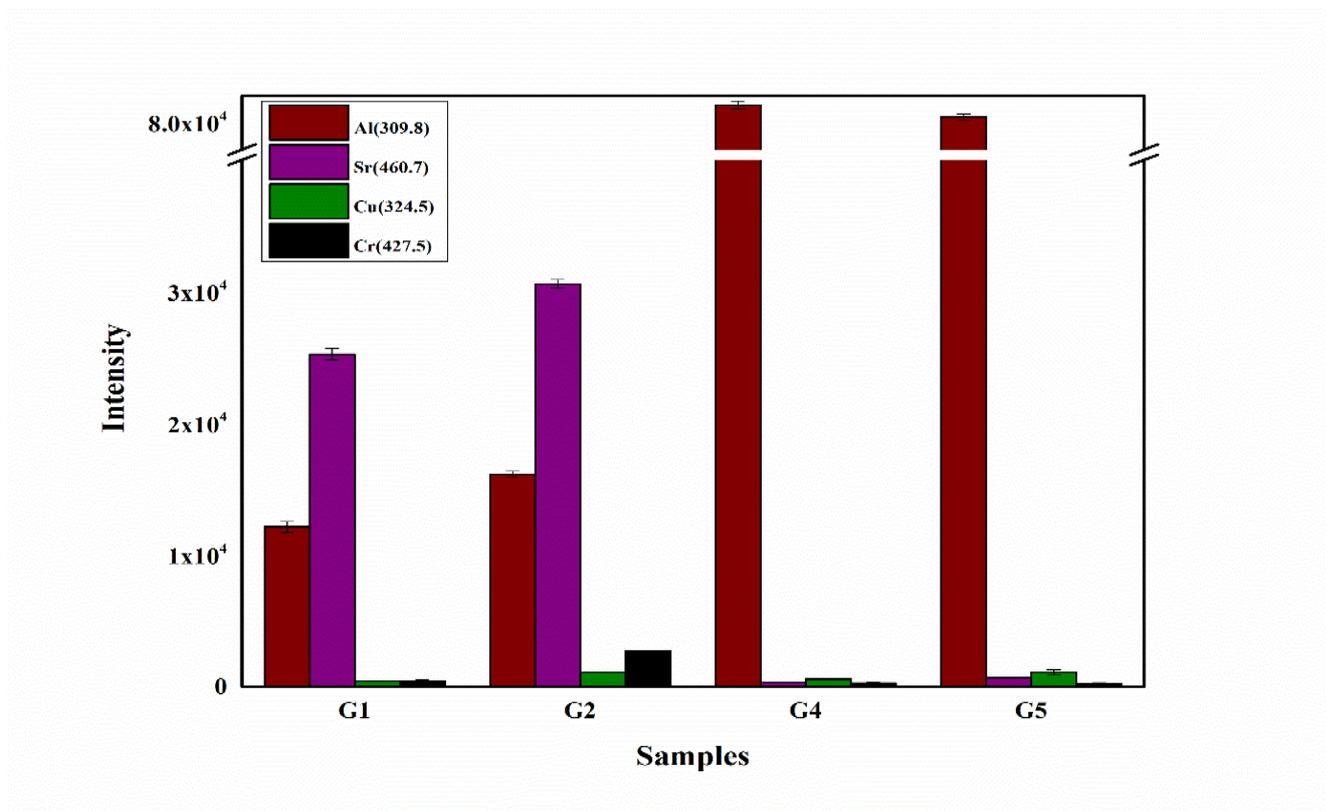

**Figure 5:** The Intensity Bar Plot ratio of the constituents Al, Cu, Cr, and Sr

It is clear from the intensity bar plat shown in Figure 5 that samples $G_1$, $G_2$ contain a higher concentration of Sr whereas $G_4$, $G_5$ contains higher concentrations of Al. The variation of concentrations in the constituents of Sample $G_3$ has the same as $G_1$, $G_2$, and $G_6$ have the same variation as $G_4$, $G_5$. Therefore, the intensity bar plot of the constituents are shown only for $G_1$, $G_2$, $G_4$, and $G_5$ in Figure 5. The results of the intensity bar plot is fairly agree with the results obtained from the qualitative analysis presented in section 3.1.

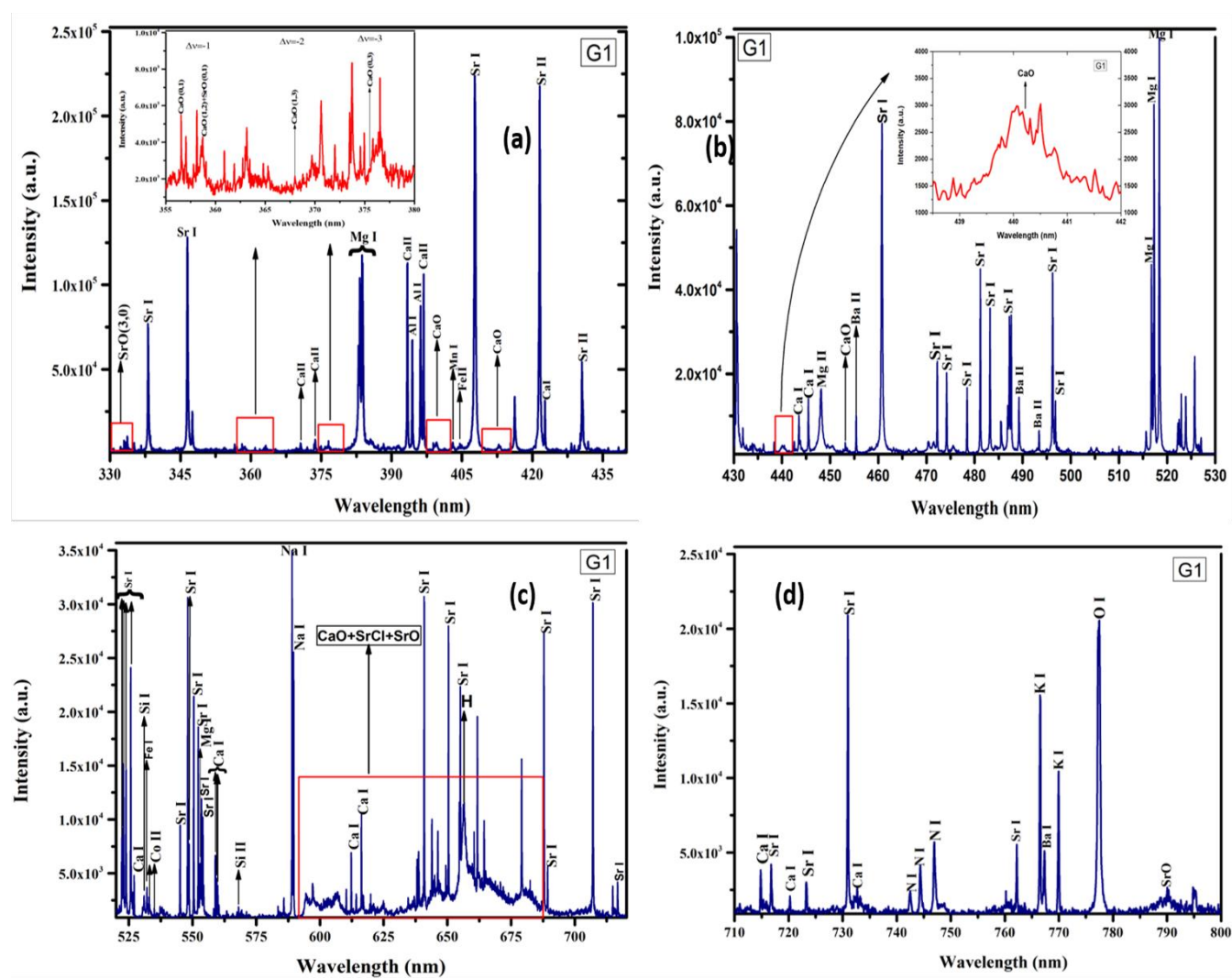

**Figure 6:** LIBS spectra of the green cracker sample $G_1$ in the spectral range 330-800nm

### 3.3. Analysis of the electronic bands of diatomic molecules

The LIBS spectra of $G_1$ shown in Figure 6 contains the spectral lines of Cu, Ca, Mg, Ba, Sr, Si, Na, H, Al, Mn, N, K, and O. The sample $G_1$ contains the higher concentration of the constituents of Sr and Ca and thus these elements react with oxygen present in the sample and form the oxide of these elements. So, the band head of CaO (0,0) (1,1), (2,2), and SrO are also found in the LIBS spectra of sample $G_1$ (CHET. R. BHATT et al., 2015; Dietz et al., 2018; Gaydon, 1964). Sample $G_1$, $G_2$, and $G_3$ contain a higher amount of constituents like Sr, Ca, Fe because these are the sparklers, and sparklers produce a lot of colours while burning. Constituents Ca is used for producing orange color, Sr used for red color, Na used for yellow color, Mg used for very bright white color and compound of Fe used for the blue and red color in green cracker sparkler also likewise normal crackers. This is the reason why group 1 contains a higher concentration of constituents

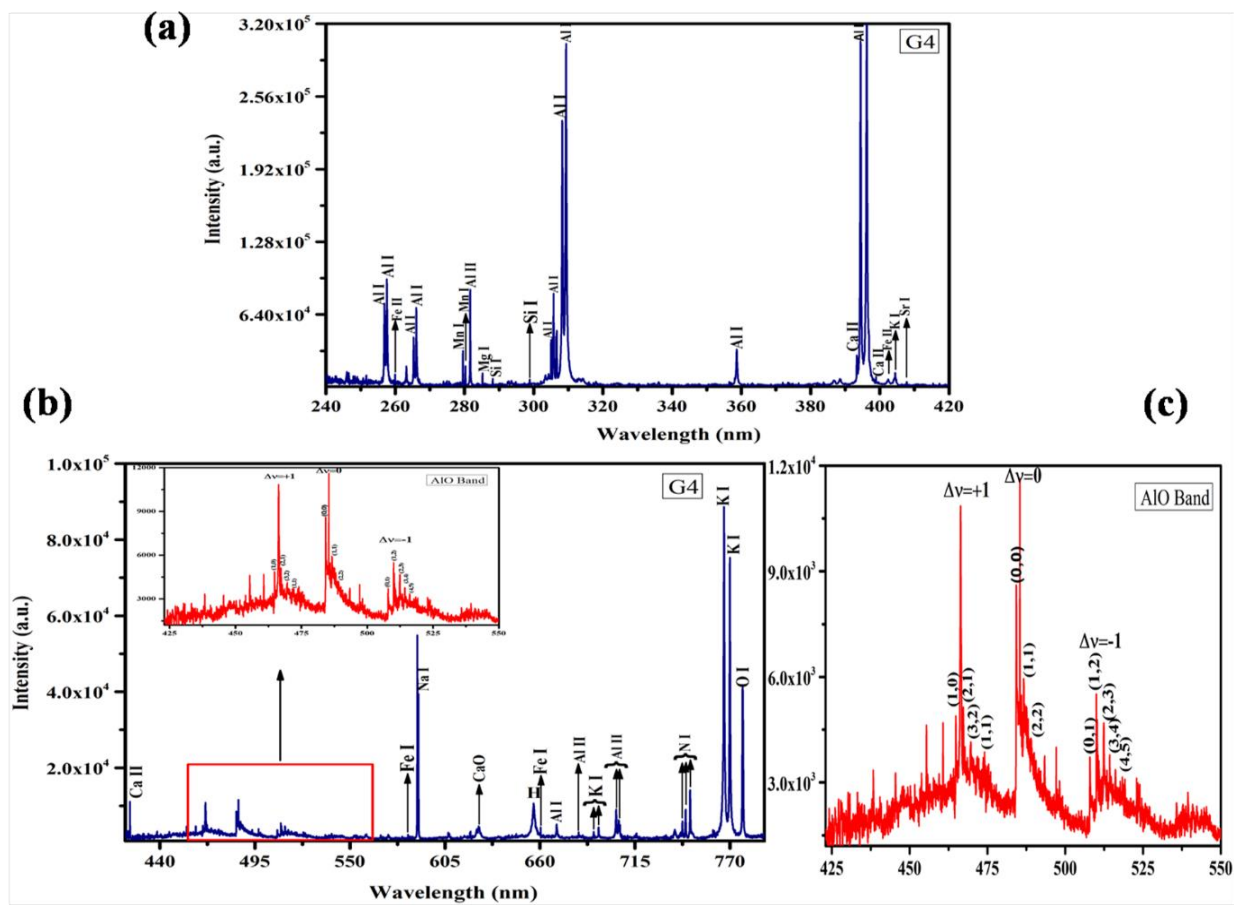

**Figure 7: a)** LIBS spectra of the sample G₄ in the spectral range of 240nm-420nm. **b)** in the spectral range of 440nm-770nm. **c)** Molecular band of AlO corresponding Δv=0, +1and -1.

LIBS spectra of sample G₄ is shown in Figure 7 which contain the spectral lines of the Na, K, O, Fe, Al, H, Ca, Mg, Si, Mn, Sr, Cr, Cu, C, and N. The sample G₄ contain the higher concentration of the Al element which react with oxygen present in the sample and form the oxide of that particular element. Therfore, the sample G₄ contains the molecular band of AlO corresponding to Δv=-1,0 and +1 which is shown in Figure 7. c). The presence of the molecular band of AlO shows that the second group (G₄, G₅, and G₆) contains a higher concentration of the Al element compared to the first group (G₁, G₂, and G₃) (Bai et al., 2014) which exactly agree with the results obtained by semiquantitative analysis discussed in section 3.2. In addition of this, the samples of the second group (G₄, G₅, and G₆) contain the higher concentration Al only. These (G$_4$, G$_5$, and G$_6$) samples are the flower pot samples and the flower pot only produces glitter and white effect after burning. The availability of higher intensity of Al element reflects that Al is used in a flower pot for the glitter and white effect.

**4. Comparison with Normal crackers:**

According to the definition of CSIR-NEERI, Green crackers contain a reduced quantity of KNO$_3$, Aluminium powder, fewer oxidizers, reduced chemical uses for colors, and substitution of BaNo$_3$ with KNO$_3$, SrNO$_3$. From the result of LIBS, it is clear that on a general scale the Sr, K are found less as said in the definition of green crackers compared to normal crackers. Apart from this, green crackers also contain a lesser amount of chemicals that produce the color. This statement is also verified because normal crackers contain more elements than green crackers like Ba etc. In addition to this, green crackers also contain a lesser amount of aluminum powder which is also true because normal crackers contain a higher Al amount (Km Darpan, 2022).

**5. Investigation of the organic compounds in firecrackers using Photoacoustic Spectroscopy (PAS technique)**

We have also used Photoacoustic Spectroscopy (PAS) which is nondestructive and based on the nonradiative transition for the identification of the molecular bands of organic and inorganic compounds. This technique is useful for scattering samples as well as for the opaque samples like firecrackers because in this technique we monitor the acoustic wave/signal induced by the absorption of electromagnetic radiation followed by nonradiative transitions. The PAS spectra of the powdered sample of G$_1$ are shown in Figure 8.a).

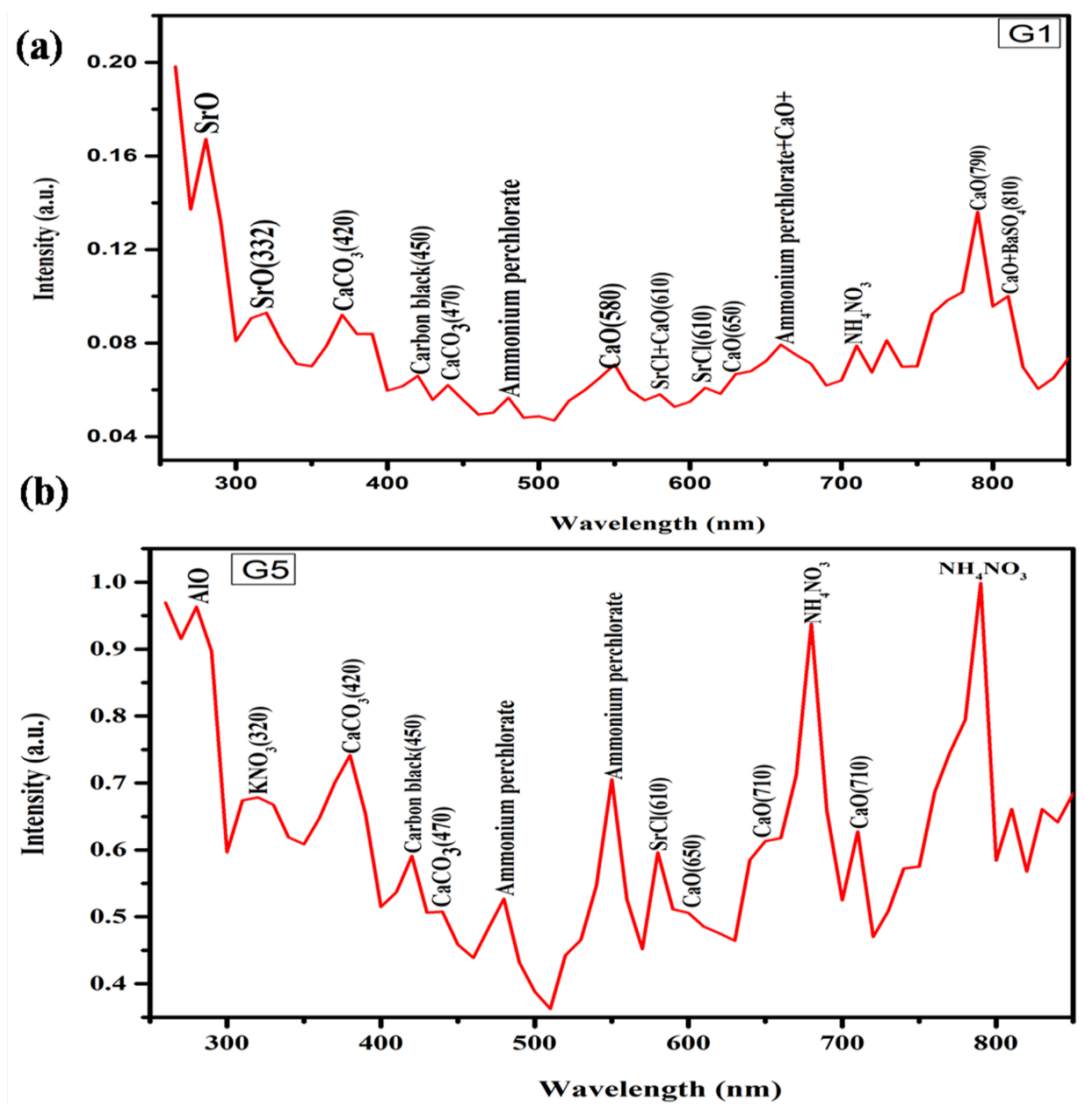

**Figure 8.a**): PAS spectra of the sample G₁ **.b**): PAS spectra of sample G₅

It is clear from Figure 8.a). that the PAS spectra of G₁ contain the molecular bands of the molecules like SrO, SrCl, CaCO₃, Charcoal, Ammonium Perchlorate, CaO, NH4NO₃, and BaSO₄ (APSANA G1, 2018; Djerdjev et al., 2018; Du & Yu, 2019; Elbasuney & El-Sharkawy, 2018; Gazalia et al., 2018; Pereira et al., 2014; Salamat Ali). The PAS spectra of the first group (G₁, G₂, and G₃) are almost the same due to which the spectra of G1 are only shown here. Similarly, the PAS spectra of the second group (G₄, G₅, and G₆) are almost the same due to which PAS spectra of G₅ are only shown in the manuscript. Molecular bands of SrO, CaO, and SrCl are present in sample $G_1$. This is also verified by the LIBS result because the LIBS spectra of sample $G_1$ contain the higher concentration of the constituents of Sr, Ca and LIBS spectra of the sample $G_1$ contain the molecular band of the SrO, CaO, and SrCl. Ammonium nitrate is found in PAS spectra of the sample $G_1$ because this ammonium nitrate is used in firecrackers for the oxidizers which helps firecrackers in burning. Ammonium perchlorate is found in PAS spectra of the sample $G_1$ because ammonium perchlorate is used in crackers because it boasts high thermal stability, high oxygen content, and low sensitivity to shock. $CaCO_3$ is used in firecrackers because when it gains temperature it decomposes into two components which is CaO and $CO_2$. When this $CaCO_3$ compound break into these compound heat produces in the firecrackers which help in burning to the firecrackers due to which the CaO molecular band is also observed in LIBS spectra as well as in PAS spectra. C is used in firecrackers in gunpowder. Gunpowder is a mixture of the sulphate group, nitrate group (potassium nitrate), sulphur, and charcoal. Due to this the PAS spectra of the sample $G_1$ contain the molecular band of $BaSO_4$ and charcoal. SrCl and SrO are used in firecrackers sample as a colouring agent, it produces the bright red color in firecrackers. Sample $G_1$ contains the Molecular band of the SrO, SrCl, and CaO because the first group ($G_1$, $G_2$, and $G_3$) contains a higher concentration of Sr and Ca in comparison to the second group ($G_4$, $G_5$ and $G_6$).

PAS spectra of sample $G_4$ are shown in Figure 6 which contain the molecular band of the AlO, KNO3, CaCO3, CaO, Charcoal, Ammonium perchlorate (K. S. Krishnan; Mohammad Mahdi Najafpour*a, 2012). The second group ($G_4$, $G_5$, and $G_6$) contains the Al element higher compared to $G_1$, $G_2$, and G3which react with the atmospheric oxygen and form the oxide of Al. Due to this the molecular band AlO is observed in PAS spectra of $G_4$. The result of the PAS technique also agrees with results of LIBS thechnique. The molecular form of CaO, CaCO3, Ammonium perchlorate, KNO3, and charcoal are mixed in green crackers for the same reason that is mixed in normal crackers..

**6. UV-Vis Spectra of Green crackers:**

To confirm the molecular bands/compositions obtained with LIBS and PAS technique, the UV-Vis spectra of all six green crackers are recorded. Since samples $G_1$, $G_2$ and $G_3$ are having almost similar compostion, the UV-Vis spectra of all three crackers are almost the same in appearance. On the other hand, the samples $G_4$, $G_5$, and $G_6$ are having almost the same character due to which the UV-Vis spectra of the samples $G_4$, $G_5$, and $G_6$ are almost the same in appearance. Tus, the UV-Vis spectra of samples G1 and G5 are shown here which corresponds to groups first and second respectively.

The UV-Vis spectra of sample $G_1$ are shown in Figure 9.a). contain the Molecular band of $CaCO_3$ and SrO. This result is according to the obtained by LIBS and PAS technique in which we observed the molecular bands of SrO and CaO due to higher concentrations of Sr and Ca in group first molecules (APSANA G1, 2018; Ghadam, 2013). Since sample $G_1$, $G_2$ and $G_3$ contain the higher intensities of the spectral lines of Sr and Ca compared to $G_4$, $G_5$, and $G_6$ which is clear from the LIBS stack plot which implies these samples contains the molecular band of SrO and CaO which is clear from LIBS and PAS spectra. Here UV-Vis spectra containing the molecular band of $CaCO_3$ and SrO are in good agreement with LIBS and PAS results. Similarly, the UV-Vis spectra of sample $G_5$ contains the molecular band of AlO and $CaCO_3$ because the second group which contains $G_4$, $G_5$, and $G_6$ samples are having a higher intensity of Al and moderate Ca element respectively (Prashanth et al., 2018). Thus UV-Vis spectra of sample $G_5$ contain a molecular band of AlO and $CaCO_3$. UV-Vis spectra of sample $G_4$ is shown in Figure.9.b) which support the results of the LIBS and PAS technique.

**(a)** **(b)**

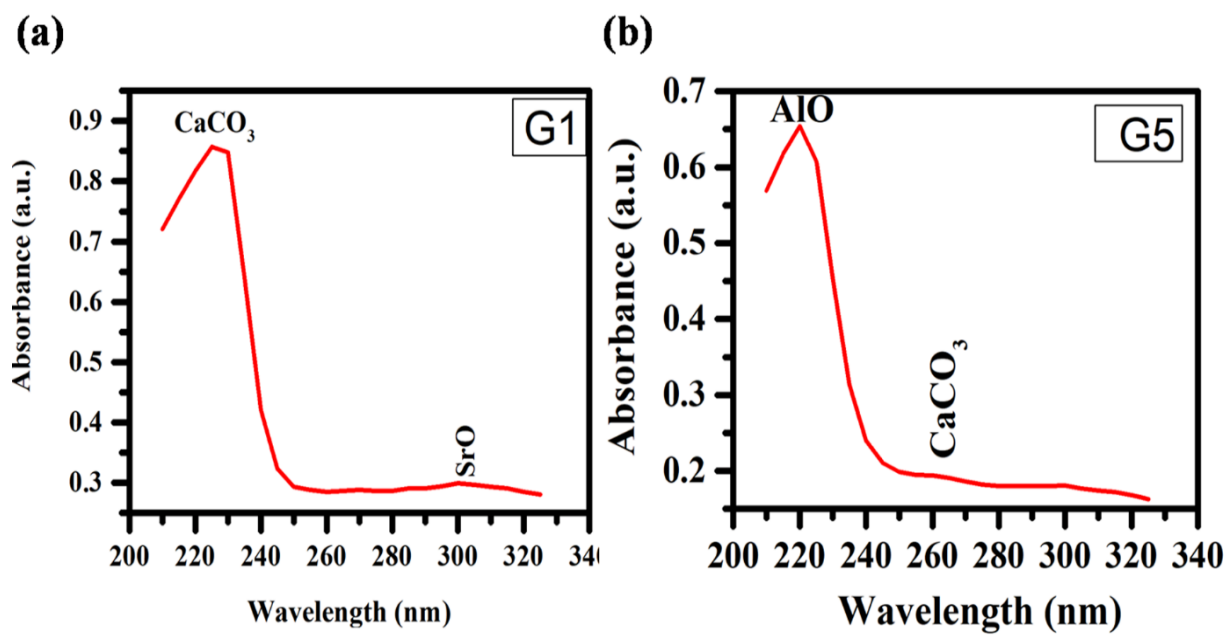

**Figure 9.a):** UV-Vis Spectra of sample G₁ **b):** UV-Visible spectra of sample G5

## 7. Evaluation of concentration of constituents using Atomic Absorption Spectroscopy (AAS):

Although the LIBS spectra of the green crackers contain many toxic elements like Al, Ba, Cr, Cu, Sr which implies that like normal crackers, green crackers are also harmful to human health. In addition to this, the observation of the molecular bands of toxic elements like Al and Sr indicates that the concentration of these elements is higher in their respective groups. But to know the actual concentration of the most toxic elements like Cr, Al, and Cu, we have used Atomic Absorption Spectroscopy (AAS) Method. Actual. The concentration of three elements Chromium (Cr), Aluminum (Al), and Copper (Cu) are tabulated in Table 4.

**Table 4:** Concentration of toxic constituents (Cr, Cu, and Al) obtained from AAS in ppm(mg/L)

| Elements | $G_1$ | $G_2$ | $G_4$ | $G_5$ |
|----------|-------|-------|-------|-------|
| Cr | 71.97 | 2598.50 | 333.98 | 204.73 |
| Al | 12377.72 | 51800.27 | 118161.21 | 94369.80 |
| Cu | 153.76 | 546.30 | 69.60 | 389.74 |

Table 4 shows that green crackers also contain a moderate amount of Cu, Cr, and Al compared to normal crackers. Therefore, the statement of  CSIR-NEERI for green crackers that contains lesser chemical constituents for color and less aluminum powder compared to normal crackers, is justified.

## 8. Principal Component Analysis (PCA):

Principal Component Analysis (PCA) is used here to classify/differentiate the six green cracker samples. PCA is applied along with LIBS data set which classifies the six samples into two groups. For the application of PCA, the dataset (278721x89) is taken and total spectra of 90 for each sample's 15 spectra are taken. A software names Unscramblers X (Camo Ltd) is used for the application of the PCA. Both, the score and loading plot of the green cracker's samples are shown in Figure .10.a) 10.b). In the Score plot (Figure 10.a), two different groups are formed. In the first group, $G_1$, $G_2$, and $G_3$ are clustered together on the positive side of the PC-1 axis, and in the second group $G_4$, $G_5$ and $G_6$ are clustered together on the negative side. Sr which is negatively correlated is present higher in intensity in $G_1$, $G_2$, and $G_3$. Al is positively correlated and higher in intensity in $G_4$, $G_5$, and $G_6$. The presence of these two constituents in different concentrations is responsible for the positive and negative correlation of the groups of the PC-1 axis. The results of the PCA exactly agree with the results obtained by LIBS, PAS, and UV-VIS spectroscopic technique.

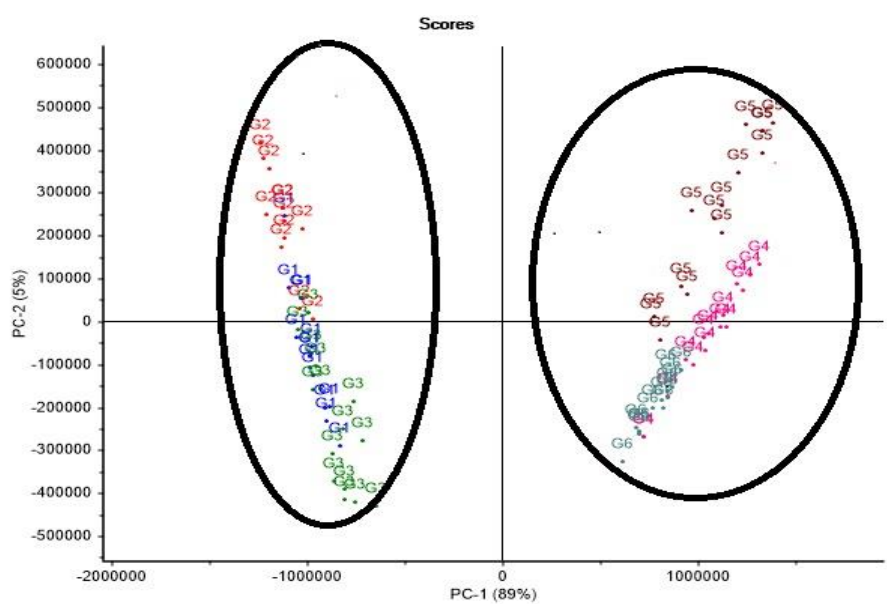

**Figure 10.a):** Score plot of six types of green crackers discriminated into two groups

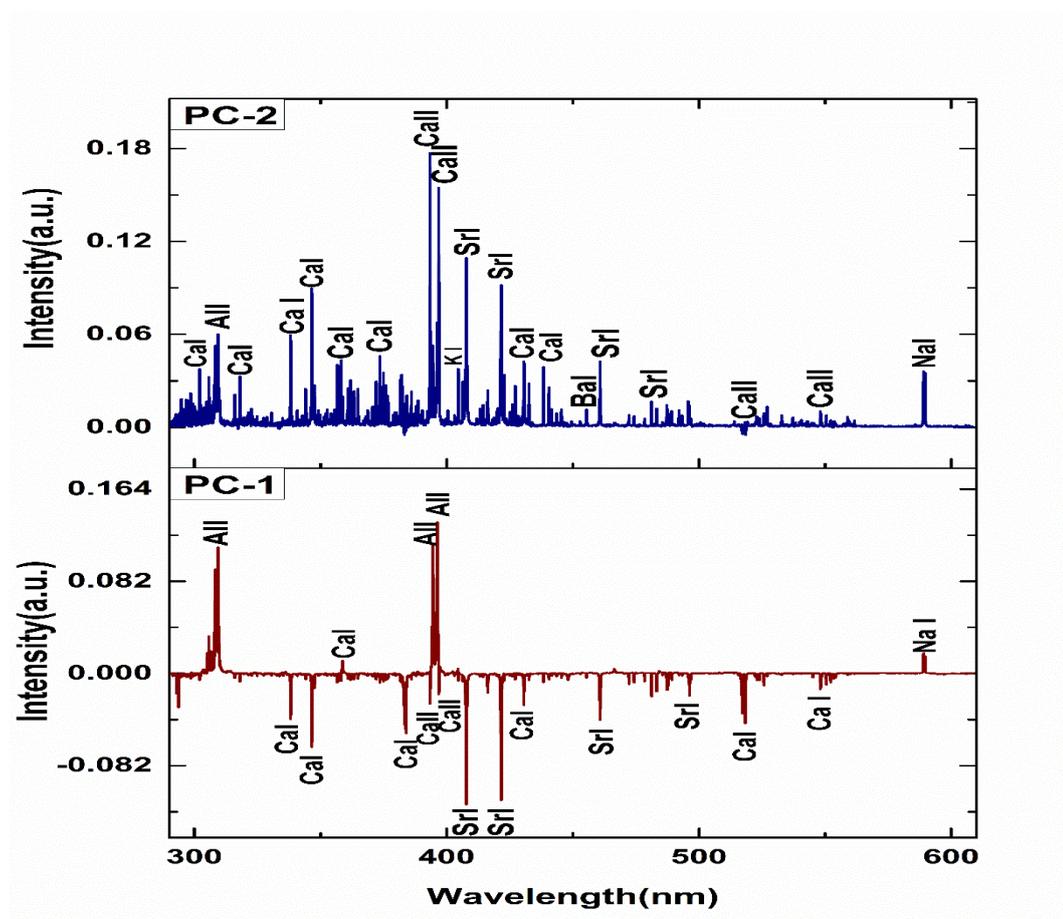

**Figure 10.b):** Loading plot of six types of green crackers

Samples G$_1$, G$_{2,}$ and G$_3$ have the same properties because these all are forming one group, and samples G$_4$, G$_{5,}$ and G$_6$ have the same properties because these are forming another group. According to PC-2 (Figure No.8(b) samples G1, G2 and G3 are differentiated into two groups in which G5 is positively correlated and G6 is negatively correlated while G4 is lie on axis. Sample G4 have the common properties of G4 and G5. Similarly second group also differentiated by PC-2, G2 is positively correlated and G3 negatively while sample G2 lie on axis because of having common properties of G2 and G3. PC-1 is unable to explain the above explanation of differentiating two groups.

According to the loading plot (Figure 10.b), samples G$_1$, G$_2$ and G$_3$ are positively correlated because these samples contain the higher intensity of Ca and Sr element which are negatively correlated in the loading plot. Samples $G_4$, $G_5$, and $G_6$ are contains the higher intensity of the Al element which positively correlated in the loading plot of PC-1 (Figure 10.b). These are the reason for differentiating these two groups. With the loading plot of PC-2, Ca is positively correlated.

## 9. Conclusions:

LIBS is a novel, quick, cheaper, and simple laser-based technique for the characterization/investigation of the lighter, heavy/toxic and minor as well as major elements/constituents present in the green crackers. Similar to traditional/normal crackers, green crackers also contain toxic elements like Al, Cr, Sr, Cu. Like normal crackers, Green crackers contain different toxic molecular bands like SrO, AlO, and nontoxic molecular bands like CaO. The presence of molecular compounds like AlO, SrO is also confirmed by PAS, and the UV-Vis technique. The presence of other organic and inorganic compounds like $CaCO_3$, $KNO_3$, $NH_4NO_3$, $NH_4ClO_4$ have been confirmed detecting their absorption spectra using PAS technique. The results of the present study reveal that PA spectra contain a large number of peaks/bands corresponding to the absorption of the various compounds/molecules in comparison to the number of peaks/bands observed in UV-VIS spectra. Thus PAS is more suitable to investigate organic compounds present in firecrackers than the UV-VIS technique. To compare the toxicity of the green crackers with traditional/normal crackers the concentrations of the toxic elements Al, Cr, Cu in green crackers are calculated using the AAS method. PCA of the LIBS is useful to discriminate/differentiate the more toxic crackers.

**Statement of environmental application:** Since Normal crackers are produce a lot of noise and pollution along with toxic gases containing SO2 & NO2. Green crackers are manufactured by CSIR-NEERI to emit less fogg and sound comparative to normal/traditional crackers.

Although green crackers are also toxic (very less comperative to normal crackers) so we can use green crackers for celibrating our festivals/big events for limited time.

**Acknowledgments:** One of the author Miss Darpan is grateful to Department of Science and Technology (DST) Inspire Fellowship for the financial support.